\documentclass[12pt,a4paper]{article}
\usepackage{epsf}
\usepackage{graphicx}
\usepackage{natbib}
\usepackage{amsmath,amssymb}
\usepackage{url}
\usepackage{cite}
\usepackage{xcolor}

\usepackage{journals_joge}

\def\fracd#1#2{\frac{\displaystyle #1}{\displaystyle #2}}

\def\ve#1{\mbox{\boldmath$#1$}}

\begin{document}

\title{On the Use of Vector Spherical Harmonics}

\author{Zinovy Malkin \\ Pulkovo Observatory, St. Petersburg 196140, Russia \\ e-mail: malkin@gaoran.ru}

\date{\vskip -2em}

\maketitle

\begin{abstract}
In many astronomical works, the structure of vector fields is analyzed, such as
the differences in the celestial object coordinates in catalogs or the celestial
object velocities, by decomposition into vector spherical harmonics (VSH).
This method has shown high efficiency in many studies, but, at the same time,
comparing the results obtained by different authors can cause difficulties
associated with different approaches to building the VSH system and even
with their different designations.
To facilitate this task, this paper provides a comparison of the three VSH
systems most often used in works on astrometry and stellar astronomy.
\end{abstract}

\section{Introduction}

Vector spherical harmonics (VSH) are widely used in astronomical
research to analyze vector fields on the celestial sphere.
In particular, their active use in astrometry began in the 1990s
with the publication of \citet{Mignard1990}.
Historically, VSH expansions in astrometry and related fields
have been used in the analysis and comparison of catalogs of stellar
positions and velocities, and in the analysis of stellar motions
to study the kinematics and dynamics of the Galaxy.

For a large number of objects, pixelization can be applied \citep{Vityazev2017B}.
In this case, the VSH decomposition is applied not to individual
stars, but to the average data over a certain set (grid) of cells
(pixels) on the celestial sphere.
Examples include the HEALPix method \citep{Gorski2005}, which
is widely used in astronomy, or the newer SREAG method
\citep{Malkin2019,Malkin2020}.
An additional advantage of the pixelization procedure is a more even
distribution of averaged data over the celestial sphere compared
to single objects.
If the pixelization is applied, the data in the cell can not only be averaged,
but also median filtering can be applied to it, which can
significantly reduce the impact of anomalous or erroneous data
(outliers) on the final result \citep{Malkin2021}.

An important circumstance for the use of VSH is the fact that
different authors implement slightly different variants of VSH
decomposition and use different harmonic notations.
The results obtained in each specific work correctly describe
the physics of the phenomenon under study, for example,
the systematic differences between the catalogs or the rotation
parameters of the Galaxy.
But the comparison of the results obtained by different authors,
in view of these differences, is not always an obvious task.
To facilitate it, this paper considers the mutual correspondence
between the VSH expansion coefficients for different implementations
of this method.


\section{Variants of VSH decompositions and their comparison}
\label{sect:vsf}

The general formula for VSH is given by the expression
\citep{Vityazev2004,Makarov2007,Mignard2012}:
\begin{equation}
\begin{array}{rcl}
\ve{T}_{lm}(\alpha,\delta) &=& \fracd{1}{\sqrt{l(l+1)}}\, \left[\fracd{\partial Y_{lm}}{\partial\delta}\,\ve{e}_\alpha -
  \fracd{1}{\cos\delta}\fracd{\partial Y_{lm}}{\partial\alpha}\,\ve{e}_\delta \right] \,, \\[2em]
\ve{S}_{lm}(\alpha,\delta) &=& \fracd{1}{\sqrt{l(l+1)}} \left[\fracd{1}{\cos\delta}\,
  \fracd{\partial Y_{lm}}{\partial\alpha}\,\ve{e}_\alpha +
  \fracd{\partial Y_{lm}}{\partial\delta}\,\ve{e}_\delta \right] \,,
\end{array}
\label{eq:vsh}
\end{equation}
where
\begin{equation*}
\ve{u}=\begin{pmatrix} \cos\alpha\cos\delta\\ \sin\alpha\cos\delta\\ \sin\delta \end{pmatrix}, \quad
\ve{e}_\alpha=\fracd{1}{\cos\delta}\,\fracd{\partial \ve{u}}{\partial\alpha}
=\begin{pmatrix} -\sin\alpha\\ \phantom{-}\cos\alpha\\ 0 \end{pmatrix}, \quad
\ve{e}_\delta=\fracd{\partial \ve{u}}{\partial\delta}
=\begin{pmatrix} -\cos\alpha\sin\delta\\ -\sin\alpha\,\sin\delta\\ \cos\delta \end{pmatrix},
\label{eq:uee}
\end{equation*}
$\ve{T}_{lm}$ and $\ve{S}_{lm}$ are toroidal and spheroidal
harmonics (often referred to in the literature as magnetic
and electrical harmonics, respectively) of degree $l$
and order $m$, $Y_{lm}$ are scalar spherical functions
of polynomials and associated Legendre functions.
Both the original spherical harmonics $Y_{lm}$ and VSH form
sets of orthonormal basis harmonics defined on a sphere.

This general formulation allows for various approaches
to its practical implementation.
It is possible to construct a set of VSH as real or
complex functions, apply various normalization methods,
and use different sign conventions.
There is not even a common approach to terminology.
For example, \citet{Mignard2012} consider the terms
spheroidal/toroidal harmonics and magnetic/electrical
harmonics to be synonymous, i.e., defining the same
harmonics under different names, whereas \citet{Vityazev2017B}
considers them different and derives different mathematical
expressions for them.

However, from the point of view of using the VSH technique
for scientific analysis, these differences in the practical
implementation of the method are not of fundamental importance,
they should be known and taken into account only for a correct
comparison of the results of different authors.
In any case, each term of the VSH decomposition consists
of a trigonometric function of the spherical coordinates
of the object and a normalizing factor.
In this paper, three most commonly used independent
approaches in astrometry and stellar astronomy are considered:
Mignard and Klioner \citep{Mignard1990,Mignard2012},
hereinafter referred to as MK, Vityazev
\citep{Vityazev2004,Vityazev2017B}, hereinafter referred to as VV
and Makarov \citep{Makarov2007}, hereinafter referred to as VM.
For the sake of completeness, other implementations of VSH
expansions can be mentioned, such as \citet{Gwinn1997,Marco2015},
but they are rarely used and can easily be added to the comparison
by an interested reader.

In practice, the problem of decomposition of the vector field
$\ve{V}(\alpha,\delta)$ under study into VSH consists of
determining the coefficients $a_i$ from the following system
of equations by the method of least squares (LS):
\begin{equation}
\ve{V}(\alpha_j,\delta_j) = \sum_{i=1}^{N_{coeff}} a_i \ve{F}_i(\alpha_j,\delta_j) \,,
\label{eq:lsp}
\end{equation}
where j is the number of the object (star, source) or cell
in the case of using pixelization,
$\ve{F}$is the generalized designation of the VSH,
$N_{coeff}$ is the number of decomposition coefficients,
which depends on the maximum degree of decomposition $l_{max}$.
For $l_{max}>0$ $N_{coeff}$=$2\,l_{max}(l_{max}+2)$,
for $l_{max}$=0 $N_{coeff}$=3 (rotation only).
Table.~\ref{tab:lmax_ncoeff} shows the $N_{coeff}$ values for $l_{max} \le 10$.

\begin{table}[ht]
\centering
\caption{The number of VSH coefficients $N_{coeff}$ depending on $l_{max}$.}
\label{tab:lmax_ncoeff}
\begin{tabular}{lccccccccccc}
\hline
$l_{max}$   & 0 & 1 &  2 &  3 &  4 &  5 &  6 &  7  &  8  &  9  & 10  \\
$N_{coeff}$ & 3 & 6 & 16 & 30 & 48 & 70 & 96 & 126 & 160 & 198 & 240 \\
\hline
\end{tabular}
\end{table}

In recent years, VSH decomposition has been widely used in
radio astrometry to analyze systematic differences between the
catalogs of the coordinates and velocities of radio sources.
For this purpose, decomposition up to $l$=2, is most often used,
as, for example, in
\citet{Liu2018,Karbon2019,Liu2020,Charlot2020,Titov2013,Yao2024}:

\begin{equation}
\begin{array}{rcl}
\Delta\alpha^{\ast} & = & R_1 \cos\alpha \sin\delta + R_2 \sin\alpha \sin\delta - R_3 \cos\delta
                    - G_1 \sin\alpha + G_2 \cos\alpha \\[1ex]
                    &   & + \, M_{2,0} \sin 2\delta
                    - (M_{2,1}^{\rm Re} \cos\alpha - M_{2,1}^{\rm Im} \sin\alpha) \cos 2\delta
                    + (E_{2,1}^{\rm Re} \sin\alpha + E_{2,1}^{\rm Im} \cos\alpha) \sin\delta \\[1.5ex]
                    &   &- \, (M_{2,2}^{\rm Re} \cos 2\alpha - M_{2,2}^{\rm Im} \sin 2\alpha) \sin 2\delta  - 2 (E_{2,2}^{\rm Re} \sin 2\alpha + E_{2,2}^{\rm Im} \cos 2\alpha) \cos\delta, \\[3ex]
\Delta\delta        & = &- \, R_1 \sin\alpha + R_2 \cos\alpha
                    - G_1 \cos\alpha \sin\delta - G_2 \sin\alpha \sin\delta + G_3 \cos\delta \\[1ex]
                    &   &+ \, E_{2,0} \sin 2\delta
                    - (M_{2,1}^{\rm Re} \sin\alpha + M_{2,1}^{\rm Im} \cos\alpha) \sin\delta
                    - \, (E_{2,1}^{\rm Re} \cos\alpha - E_{2,1}^{\rm Im} \sin\alpha) \cos 2\delta \\[1.5ex]
                    &   &+ \, 2 (M_{2,2}^{\rm Re} \sin 2\alpha + M_{2,2}^{\rm Im} \cos 2\alpha) \cos\delta
                    - \, (E_{2,2}^{\rm Re} \cos 2\alpha - E_{2,2}^{\rm Im} \sin 2\alpha) \sin 2\delta, \\[1ex]
\end{array}
\label{eq:vsh2}
\end{equation}
where $\Delta\alpha^* = \Delta\alpha\cos\delta$,
$\Delta\alpha$ and $\Delta\delta$ are the differences
in the coordinates of the object in the compared catalogs.

As can be seen from this example, in the practice of using VSH,
the coefficients of the model are often determined directly
with trigonometric harmonics without taking into account
the normalizing factor.
Such an approach can often be justified by a direct correspondence
between the found coefficients and the parameters of a physical
phenomenon, for example, the mutual rotation between catalogs
or the kinematic parameters of the Galaxy.

Six harmonics of the first degree form the rotation vector
$\mathbf{R} = (R_1, R_2, R_3)^{\mathrm{T}}$
and the dipole deformation vector
$\mathbf{G} = (G_1, G_2, G_3)^{\mathrm{T}}$,
called \textit{glide} by \citet{Mignard2012},
and \textit{secular aberration drift} by \citet{Titov2013},
which is associated with phenomena such as Galactic aberration
in proper motion, caused by the centripetal acceleration of
the Solar System as it rotates around the center of the Galaxy.
Note that in expression~(\ref{eq:vsh2}), the signs of the
components of the rotation vector correspond to the classical
definition of the direction of rotation between catalogs
\citep{Walter2000}:
\begin{equation}
\begin{array}{rcl}
\Delta\alpha^{\ast} & = & \phantom{-}R_1\cos\alpha\sin\delta + R_2\sin\alpha\sin\delta - R_3\cos\delta \,, \\[1ex]
\Delta\delta & = & -R_1\sin\alpha + R_2\cos\alpha \,.
\end{array}
\label{eq:rotation_tanDel}
\end{equation}

In some papers, for example \citep{Liu2018,Liu2020},
the components of the vectors $\mathbf{R}$ and $\mathbf{G}$
are given with the opposite sign, but this presents no
difficulty in comparing the results, although it should
be tracked and taken into account.

In Table~\ref{tab:vsh2}, expressions (\ref{eq:vsh2}) are
given as a set of harmonics before which the coefficients
are directly fitted by the LS (see Eq.~\ref{eq:lsp})
if the calculations are performed without taking into
account the normalizing factors.

\begin{table}[ht]
\centering
\caption{VSH functions up to $l_{max}$=2, corresponding to expression (\ref{eq:vsh2}).}
\label{tab:vsh2}
\begin{tabular}{ccc}
\hline
Term & $\Delta\alpha^{\ast}$ & $\Delta\delta$ \\
\hline
$R_1$              & $\cos\alpha \sin\delta$    & $-\sin\alpha$              \\[1ex]
$R_2$              & $\sin\alpha \sin\delta$    & $\cos\alpha$               \\[1ex]
$R_3$              & $-\cos\delta$              & ---                        \\[1ex]
$G_1$              & $-\sin\alpha$              & $-\cos\alpha \sin\delta$   \\[1ex]
$G_2$              & $\cos\alpha$               & $-\sin\alpha \sin\delta$   \\[1ex]
$G_3$              & ---                        & $\cos\delta$               \\[1ex]
$M_{2,0}$          & $\sin2\delta$              & ---                        \\[1ex]
$E_{2,0}$          & ---                        & $\sin2\delta$              \\[1.5ex]
$M_{2,1}^{\rm Re}$ & $-\cos\alpha \cos2\delta$  & $-\sin\alpha \sin\delta$   \\[1.5ex]
$M_{2,1}^{\rm Im}$ & $\sin\alpha \cos2\delta$   & $-\cos\alpha \sin\delta$   \\[1.5ex]
$E_{2,1}^{\rm Re}$ & $\sin\alpha \sin\delta$    & $-\cos\alpha \cos2\delta$  \\[1.5ex]
$E_{2,1}^{\rm Im}$ & $\cos\alpha \sin\delta$    & $\sin\alpha \cos2\delta$   \\[1.5ex]
$M_{2,2}^{\rm Re}$ & $-\cos2\alpha \sin2\delta$ & $2\sin2\alpha \cos\delta$  \\[1.5ex]
$M_{2,2}^{\rm Im}$ & $\sin2\alpha \sin2\delta$  & $2\cos2\alpha \cos\delta$  \\[1.5ex]
$E_{2,2}^{\rm Re}$ & $-2\sin2\alpha \cos\delta$ & $-\cos2\alpha \sin2\delta$ \\[1.5ex]
$E_{2,2}^{\rm Im}$ & $-2\cos2\alpha \cos\delta$ & $\sin2\alpha \sin2\delta$  \\[1ex]
\hline
\end{tabular}
\end{table}

Explicit expressions for higher degrees of VSH were published
by \citet{Mignard2012} (MK decomposition, $l_{max}$=4)
and \citet{Vityazev2017B} (VV decomposition, $l_{max}$=3).
Expressions for the VM decomposition up to $l_{max}$=4 were
kindly provided by their author \citep{Makarov2024}.
Table~\ref{tab:vsh4_t} shows 24 toroidal harmonics, and
Table~\ref{tab:vsh4_s} shows 24 spheroidal harmonics,
together forming a complete set of 48 functions for $l_{max}$=4
for the three compared VSH expansions.
It should be noted that in the MK decomposition all
normalizing factors are positive, and in the VV and VM
expansions, the set of normalizing factors is alternating.
In the last two cases, the sign was transferred
to the trigonometric function.
Therefore, the signs in Table~\ref{tab:vsh4_t} and
\ref{tab:vsh4_s} correspond to the case of determining
the coefficients of VSH decomposition without taking into
account the normalizing factors.
From the data in these tables, it can be seen that the signs
of the same harmonics in VM and MK are the same for even
values of $l$+$m$, and the signs of the same harmonics
in VV and MK are the same for even $m$.

Let us separately describe the structure of VSH notations
for the three expansions under consideration.
The MK decomposition in the original work is presented
in a complex form, so the designation of the harmonics
of this expansion in this work is
$\{\ve{T}|\ve{S}\}_{lm}^{\{Re|Im\}}$,
where $\ve{T}$ and $\ve{S}$ stand for toroidal and
spheroidal harmonics, respectively, and the superscripts
$Re$ and $Im$ mean real and imaginary components of
harmonics, respectively.
The decompositions VV and VM are made in real functions
and are represented in the present work in their original
notations.
The VV designations are similar to the MK designations,
with the only difference that instead of the index of
the real or imaginary part, a third sub-index is used
(1 corresponds to the real part and 0 to the imaginary part).
The notation of harmonics in the VM decomposition consists
of four indices and has the form $(\{mag|ele\},k,l,m)$,
where $mag$ and $ele$ stand for toroidal and spheroidal
harmonics, respectively, and $k$ has a value of 0 at $m=0$,
and at $m>0$ $k$=1 corresponds to the real part and
$k$ = 2 to the imaginary part.

\begin{table}[p]
\centering
\caption{Toroidal harmonics up to $l_{max}$=4.
  L2~-- expression (\ref{eq:vsh2}), MK~-- \citet{Mignard2012},
  VV~-- \citet{Vityazev2017B}, VM~-- \citet{Makarov2024}.}
\label{tab:vsh4_t}
\tabcolsep=4pt
\begin{tabular}{cccccc}
\hline
MK & L2 & VV & VM & $\alpha^{\ast} $& $\delta $\\
\hline \\[-6pt]
$\ve{T}_{10}$          & $-R_3$              & $~\ve{T}_{1,0,1}$ & $-$(mag,0,1,0) & $\cos\delta$                                      & ---                                         \\[1ex]
$\ve{T}_{11}^{\rm Re}$ & $~R_1$              & $-\ve{T}_{1,1,1}$ & \  (mag,1,1,1) & $\sin\delta\cos\alpha$                            & $-\sin\alpha$                               \\[1ex]
$\ve{T}_{11}^{\rm Im}$ & $~R_2$              & $-\ve{T}_{1,1,0}$ & \  (mag,2,1,1) & $\sin\delta\sin\alpha$                            & $ \cos\alpha$                               \\[1ex]
$\ve{T}_{20}$          & $M_{2,0}$           & $~\ve{T}_{2,0,1}$ & \  (mag,0,2,0) & $\sin2\delta$                                     & ---                                         \\[1ex]
$\ve{T}_{21}^{\rm Re}$ & $~M_{2,1}^{\rm Re}$ & $-\ve{T}_{2,1,1}$ & $-$(mag,1,2,1) & $-\cos2\delta\cos\alpha$                          & $-\sin\delta\sin\alpha$                     \\[1ex]
$\ve{T}_{21}^{\rm Im}$ & $-M_{2,1}^{\rm Im}$ & $-\ve{T}_{2,1,0}$ & $-$(mag,2,2,1) & $-\cos2\delta\sin\alpha$                          & $ \sin\delta\cos\alpha$                     \\[1ex]
$\ve{T}_{22}^{\rm Re}$ & $~M_{2,2}^{\rm Re}$ & $~\ve{T}_{2,2,1}$ & \  (mag,1,2,2) & $-\sin2\delta\cos2\alpha$                         & $ 2\cos\delta\sin2\alpha$                   \\[1ex]
$\ve{T}_{22}^{\rm Im}$ & $-M_{2,2}^{\rm Im}$ & $~\ve{T}_{2,2,0}$ & \  (mag,2,2,2) & $-\sin2\delta\sin2\alpha$                         & $-2\cos\delta\cos2\alpha$                   \\[1ex]
$\ve{T}_{30}$          &                     & $~\ve{T}_{3,0,1}$ & $-$(mag,0,3,0) & $\cos\delta(5\sin^2\!\delta-1)$                   & ---                                         \\[1ex]
$\ve{T}_{31}^{\rm Re}$ &                     & $-\ve{T}_{3,1,1}$ & \  (mag,1,3,1) & $\sin\delta(15\sin^2\!\delta-11)\cos\alpha$       & $-(5\sin^2\!\delta-1)\sin\alpha$            \\[1ex]
$\ve{T}_{31}^{\rm Im}$ &                     & $-\ve{T}_{3,1,0}$ & \  (mag,2,3,1) & $\sin\delta(15\sin^2\!\delta-11)\sin\alpha$       & $ (5\sin^2\!\delta-1)\cos\alpha$            \\[1ex]
$\ve{T}_{32}^{\rm Re}$ &                     & $~\ve{T}_{3,2,1}$ & $-$(mag,1,3,2) & $-\cos\delta(3\sin^2\!\delta-1)\cos2\alpha$       & $ \sin2\delta\sin2\alpha$                   \\[1ex]
$\ve{T}_{32}^{\rm Im}$ &                     & $~\ve{T}_{3,2,0}$ & $-$(mag,2,3,2) & $-\cos\delta(3\sin^2\!\delta-1)\sin2\alpha$       & $-\sin2\delta\cos2\alpha$                   \\[1ex]
$\ve{T}_{33}^{\rm Re}$ &                     & $-\ve{T}_{3,3,1}$ & \  (mag,1,3,3) & $\cos^2\!\delta\sin\delta\cos3\alpha$             & $-\cos^2\!\delta\sin3\alpha$                \\[1ex]
$\ve{T}_{33}^{\rm Im}$ &                     & $-\ve{T}_{3,3,0}$ & \  (mag,2,3,3) & $\cos^2\!\delta\sin\delta\sin3\alpha$             & $ \cos^2\!\delta\cos3\alpha$                \\[1ex]
$\ve{T}_{40}$          &                     &                   & \  (mag,0,4,0) & $\sin2\delta(7\sin^2\!\delta-3)$                  & ---                                         \\[1ex]
$\ve{T}_{41}^{\rm Re}$ &                     &                   & $-$(mag,1,4,1) & $(28\sin^4\!\delta-27\sin^2\!\delta+3)\cos\alpha$ & $-\sin\delta(7\sin^2\!\delta-3)\sin\alpha$  \\[1ex]
$\ve{T}_{41}^{\rm Im}$ &                     &                   & $-$(mag,2,4,1) & $(28\sin^4\!\delta-27\sin^2\!\delta+3)\sin\alpha$ & $ \sin\delta(7\sin^2\!\delta-3)\cos\alpha$  \\[1ex]
$\ve{T}_{42}^{\rm Re}$ &                     &                   & \  (mag,1,4,2) & $-\sin2\delta(7\sin^2\!\delta-4)\cos2\alpha$      & $ \cos\delta(7\sin^2\!\delta-1)\sin2\alpha$ \\[1ex]
$\ve{T}_{42}^{\rm Im}$ &                     &                   & \  (mag,2,4,2) & $-\sin2\delta(7\sin^2\!\delta-4)\sin2\alpha$      & $-\cos\delta(7\sin^2\!\delta-1)\cos2\alpha$ \\[1ex]
$\ve{T}_{43}^{\rm Re}$ &                     &                   & $-$(mag,1,4,3) & $\cos^2\!\delta(4\sin^2\!\delta-1)\cos3\alpha$    & $-3\cos^2\!\delta\sin\delta\sin3\alpha$     \\[1ex]
$\ve{T}_{43}^{\rm Im}$ &                     &                   & $-$(mag,2,4,3) & $\cos^2\!\delta(4\sin^2\!\delta-1)\sin3\alpha$    & $ 3\cos^2\!\delta\sin\delta\cos3\alpha$     \\[1ex]
$\ve{T}_{44}^{\rm Re}$ &                     &                   & \  (mag,1,4,4) & $-\cos^3\!\delta\sin\delta\cos4\alpha$            & $ \cos^3\!\delta\sin4\alpha$                \\[1ex]
$\ve{T}_{44}^{\rm Im}$ &                     &                   & \  (mag,2,4,4) & $-\cos^3\!\delta\sin\delta\sin4\alpha$            & $-\cos^3\!\delta\cos4\alpha$                \\[1ex]
\hline
\end{tabular}
\end{table}

\begin{table}[p]
\centering
\caption{Spheroidal harmonics up to $l_{max}$=4.
  L2~-- expression (\ref{eq:vsh2}), MK~-- \citet{Mignard2012},
  VV~-- \citet{Vityazev2017B}, VM~-- \citet{Makarov2024}.}
\label{tab:vsh4_s}
\tabcolsep=4pt
\begin{tabular}{cccccc}
\hline
MK & L2 & VV & VM & $\alpha^{\ast} $& $\delta $\\
\hline \\[-6pt]
$\ve{S}_{10}$          & $~G_3$              & $~\ve{S}_{1,0,1}$ & $-$(ele,0,1,0) & ---                                         & $\cos\delta$                                      \\[1ex]
$\ve{S}_{11}^{\rm Re}$ & $-G_1$              & $-\ve{S}_{1,1,1}$ & \  (ele,1,1,1) & $\sin\alpha$                                & $\sin\delta\cos\alpha$         \\[1ex]
$\ve{S}_{11}^{\rm Im}$ & $-G_2$              & $-\ve{S}_{1,1,0}$ & \  (ele,2,1,1) & $-\cos\alpha$                               & $\sin\delta\sin\alpha$                            \\[1ex]
$\ve{S}_{20}$          & $E_{2,0}$           & $~\ve{S}_{2,0,1}$ & \  (ele,0,2,0) & ---                                         & $\sin2\delta$                                     \\[1ex]
$\ve{S}_{21}^{\rm Re}$ & $~E_{2,1}^{\rm Re}$ & $-\ve{S}_{2,1,1}$ & $-$(ele,1,2,1) & $\sin\delta\sin\alpha$                      & $-\cos2\delta\cos\alpha$                          \\[1ex]
$\ve{S}_{21}^{\rm Im}$ & $-E_{2,1}^{\rm Im}$ & $-\ve{S}_{2,1,0}$ & $-$(ele,2,2,1) & $-\sin\delta\cos\alpha$                     & $-\cos2\delta\sin\alpha$                          \\[1ex]
$\ve{S}_{22}^{\rm Re}$ & $~E_{2,2}^{\rm Re}$ & $~\ve{S}_{2,2,1}$ & \  (ele,1,2,2) & $-2\cos\delta\sin2\alpha$                   & $-\sin2\delta\cos2\alpha$                         \\[1ex]
$\ve{S}_{22}^{\rm Im}$ & $-E_{2,2}^{\rm Im}$ & $~\ve{S}_{2,2,0}$ & \  (ele,2,2,2) & $ 2\cos\delta\cos2\alpha$                   & $-\sin2\delta\sin2\alpha$                         \\[1ex]
$\ve{S}_{30}$          &                     & $~\ve{S}_{3,0,1}$ & $-$(ele,0,3,0) & ---                                         & $\cos\delta(5\sin^2\!\delta-1)$                   \\[1ex]
$\ve{S}_{31}^{\rm Re}$ &                     & $-\ve{S}_{3,1,1}$ & \  (ele,1,3,1) & $ (5\sin^2\!\delta-1)\sin\alpha$            & $\sin\delta(15\sin^2\!\delta-11)\cos\alpha$       \\[1ex]
$\ve{S}_{31}^{\rm Im}$ &                     & $-\ve{S}_{3,1,0}$ & \  (ele,2,3,1) & $-(5\sin^2\!\delta-1)\cos\alpha$            & $\sin\delta(15\sin^2\!\delta-11)\sin\alpha$       \\[1ex]
$\ve{S}_{32}^{\rm Re}$ &                     & $~\ve{S}_{3,2,1}$ & $-$(ele,1,3,2) & $-\sin2\delta\sin2\alpha$                   & $-\cos\delta(3\sin^2\!\delta-1)\cos2\alpha$       \\[1ex]
$\ve{S}_{32}^{\rm Im}$ &                     & $~\ve{S}_{3,2,0}$ & $-$(ele,2,3,2) & $ \sin2\delta\cos2\alpha$                   & $-\cos\delta(3\sin^2\!\delta-1)\sin2\alpha$       \\[1ex]
$\ve{S}_{33}^{\rm Re}$ &                     & $-\ve{S}_{3,3,1}$ & \  (ele,1,3,3) & $ \cos^2\!\delta\sin3\alpha$                & $\cos^2\!\delta\sin\delta\cos3\alpha$             \\[1ex]
$\ve{S}_{33}^{\rm Im}$ &                     & $-\ve{S}_{3,3,0}$ & \  (ele,2,3,3) & $-\cos^2\!\delta\cos3\alpha$                & $\cos^2\!\delta\sin\delta\sin3\alpha$             \\[1ex]
$\ve{S}_{40}$          &                     &                   & \  (ele,0,4,0) & ---                                         & $\sin2\delta(7\sin^2\!\delta-3)$                  \\[1ex]
$\ve{S}_{41}^{\rm Re}$ &                     &                   & $-$(ele,1,4,1) & $ \sin\delta(7\sin^2\!\delta-3)\sin\alpha$  & $(28\sin^4\!\delta-27\sin^2\!\delta+3)\cos\alpha$ \\[1ex]
$\ve{S}_{41}^{\rm Im}$ &                     &                   & $-$(ele,2,4,1) & $-\sin\delta(7\sin^2\!\delta-3)\cos\alpha$  & $(28\sin^4\!\delta-27\sin^2\!\delta+3)\sin\alpha$ \\[1ex]
$\ve{S}_{42}^{\rm Re}$ &                     &                   & \  (ele,1,4,2) & $-\cos\delta(7\sin^2\!\delta-1)\sin2\alpha$ & $-\sin2\delta(7\sin^2\!\delta-4)\cos2\alpha$      \\[1ex]
$\ve{S}_{42}^{\rm Im}$ &                     &                   & \  (ele,2,4,2) & $ \cos\delta(7\sin^2\!\delta-1)\cos2\alpha$ & $-\sin2\delta(7\sin^2\!\delta-4)\sin2\alpha$      \\[1ex]
$\ve{S}_{43}^{\rm Re}$ &                     &                   & $-$(ele,1,4,3) & $ 3\cos^2\!\delta\sin\delta\sin3\alpha$     & $\cos^2\!\delta(4\sin^2\!\delta-1)\cos3\alpha$    \\[1ex]
$\ve{S}_{43}^{\rm Im}$ &                     &                   & $-$(ele,2,4,3) & $-3\cos^2\!\delta\sin\delta\cos3\alpha$     & $\cos^2\!\delta(4\sin^2\!\delta-1)\sin3\alpha$    \\[1ex]
$\ve{S}_{44}^{\rm Re}$ &                     &                   & \  (ele,1,4,4) & $-\cos^3\!\delta\sin4\alpha$                & $-\cos^3\!\delta\sin\delta\cos4\alpha$            \\[1ex]
$\ve{S}_{44}^{\rm Im}$ &                     &                   & \  (ele,2,4,4) & $ \cos^3\!\delta\cos4\alpha$                & $-\cos^3\!\delta\sin\delta\sin4\alpha$            \\[1ex]
\hline
\end{tabular}
\end{table}


\section{Conclusions}
\label{sect:conclusion}

The VSH decomposition method provides a powerful mathematical
tool for the analysis of vector fields on the celestial sphere,
such as the proper motions of stars or the differences in the
coordinates of stars or extragalactic sources in two catalogs.
Therefore, the VSH is often used in various astronomical studies,
in particular in the field of astrometry and Galactic astronomy.
In addition to the examples given above, a number of important,
primarily astrometric, results obtained using this method
can be noted.
\citet{Vityazev2014} conducted a detailed analysis of the
star velocity field in the UCAC4, XPM and PPMXL catalogs
for refined analysis of the rotation of the Galaxy.
Systematic differences between the coordinates and velocities
of stars in these three catalogs were studied in
\citet{Vityazev2015a,Vityazev2015b}.
The joint processing of the coordinates, proper motions,
parallaxes, and radial velocities of stars from the RAVE5,
UCAC4, PPMXL, and $Gaia$ TGAS catalogs made it possible
to significantly refine the kinematic parameters of the Galaxy
\citep{Vityazev2017Ap}.
\citet{Makarov2023} studied systematic differences in the
position catalogs of extragalactic radio sources obtained
from VLBI (ICRF3, \citet{Charlot2020}) and space
($Gaia$-CRF3, \citet{Klioner2022}) observations,
and derived a combined catalog based on two methods
of observation was obtained for the first time.
Analysis of the differences between the ICRF3 and $Gaia$-CRF2
catalogs \citep{Mignard2018} was used to verify and orient
the latter.
The application of VSH to analysis of the velocity field
of radio sources obtained from VLBI observations made
it possible to refine the parameters of the Galactocentric
acceleration of the Solar System \citep{Titov2013}.
This list can be extended.

However, the practical implementation of the theory of VSH
allows for some freedom in the choice of decomposition details.
This is not essential for the scientific interpretation
of the results obtained in the course of a particular study,
but for a correct comparison of the results presented
in different publications, it is necessary to take into account
the details of the VSH decomposition variants used to obtain them,
which make it possible to determine the correspondence of
the obtained decomposition coefficients.
This work is an attempt to at least partially facilitate this task.

One of the conclusions from this work can be found in
the recommendation to provide detailed information about
the decompositions used when publishing the results obtained
using the VSH method.

Finally, it should be noted that the author did not set
himself the task of a complete comparison of all the used
(and, moreover, all possible) variants of VSH decompositions.
The main goal of this work is to draw attention to this problem.


\section*{Acknowledgments}

The author is grateful to V. Makarov for providing explicit
expressions of the harmonics of his VSH decomposition,
which are used in the work.
The author is grateful to the reviewer for useful suggestions
for improving the initial version of the article.
This research has made use of SAO/NASA Astrophysics Data System
(ADS),
\url{https://ui.adsabs.harvard.edu/},

\bibliography{malkin_vsh}
\bibliographystyle{joge}

\end{document}